# ON THE RELEVANCE OF LANGUAGE IN SPEAKER RECOGNITION[1]

*Antonio Satué-Villar, Marcos Faundez-Zanuy*
Escola Universitària Politècnica de Mataró
Universitat Politècnica de Catalunya (UPC)
Avda. Puig i Cadafalch 101-111, E-08303 Mataró (BARCELONA)
e-mail: {satue,faundez}@eupmt.es  http://www.eupmt.es/veu

## ABSTRACT

This paper presents a new database collected from a bilingual speakers set (49), in two different languages: Spanish and Catalan. Phonetically there are significative differences between both languages. These differences have let us to establish several conclusions on the relevance of language in speaker recognition, using two methods: vector quantization and covariance matrices.

## 1. INTRODUCTION

There is a considerable literature on speaker recognition with different languages and databases. We think that it is necessary to take into account the language of the database when different methods are compared (evaluated): the particular language of the database can yield to an improvement of recognition rates.

In this paper we want to study if for bilingual speakers it is possible to obtain a more robust model combining information of both languages and how.

We have done a set of experiments with a new bilingual database in order to establish if the language of the speaker has relevance in a speaker identification application (mainly if it is more suitable one language than other, and if it is possible to recognize with different training and test languages).

This paper will present a new database collected from a bilingual speakers set (49), in two different languages: Spanish and Catalan. Phonetically there are significative differences between both languages. Mainly, the Catalan language has eight vowels (see figure 1) and Spanish only five. Although there are only nine million people of Catalan speakers in front of four hundred million people of Spanish, both languages can be used for our purpose. The differences between both languages has let us to establish several conclusions on the relevance of language in speaker recognition.

Another important question is that for bilingual speakers in conversational speech is quite common the change from one language to the other, so it is interesting to evaluate if this fact can affect a speaker recognizer.

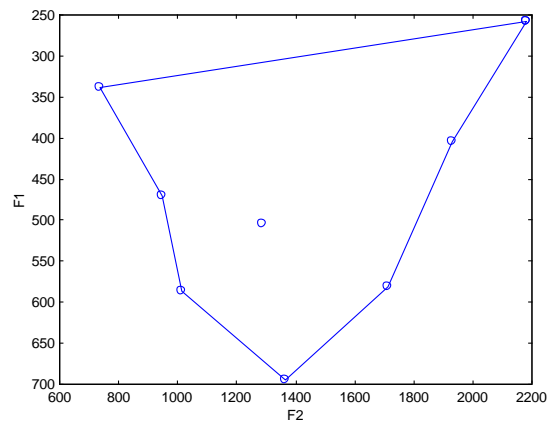

Figure 1: Formants of the Catalan vowels

## 2. NEW DATABASE

In the context of a Spanish ministry grant, and in collaboration with the Departamento de Ingeniería audiovisual y comunicaciones (DIAC) we have acquired a new database. Information about the design of the speech corpus can be found at [1]. An interesting fact is that the Spanish sentences have been balanced, but the Catalan ones have been merely translated from Spanish. Thus, the database consists on the same texts recorded in both languages in the same day, one language after the other. Speaker could freely choose which was the first recording language.

Main characteristics of the database are:

- 4 sessions with different tasks in each session (isolated numbers, connected numbers, sentences, text, conversational speech, etc.)

- In each session tasks were sequentially collected in both languages (Catalan and Spanish) uttered from the same speaker. Each task was simultaneously captured with two microphones (SONY ECM-66B and AKG C-420).

[1] This work has been supported by the CICYT TIC97-1001-C02-02

This paper presents the results of the fourth session using the common text (approx., 1 min) and the first five sentences (approx., 4 seconds each one).

## 3. SPEAKER IDENTIFICATION EXPERIMENTS

With this database we have made several tests:

- Speaker recognition with each language (train and test in Catalan, train and test in Spanish)

- Speaker recognition with different train and test conditions (train in Catalan and test in Spanish, train in Spanish and test in Catalan).

Two speaker identification methods have been used:

1. Vector quantization [2] with a random method for codebook generation (1 codebook for each speaker). The number of parameters used in each model is:

$$parameters = 2^{No} \times P$$

where P is the analysis order of the parameterization (dimension of LPCC vectors) and No is the number of bits of the codebook ranging from 0 to 7.

2. Arithmetic-harmonic sphericity measure [3], which implies the computation of a covariance matrix for each speaker, and the following measure distance:

$$\mu(C_j C_{test}) = log[tr(C_{test} C_j^{-1}) tr(C_j C_{test}^{-1})] - 2log(m)$$

where $C_i$ is a covariance matrix and m is its size.

The trace of the matrices can be computed as:

$$tr(YX^{-1}) = 2 \sum_{i=1}^{m} \sum_{j=1}^{i-1} y_{ij} \tilde{x}_{ij} + \sum_{k=1}^{m} y_{kk} \tilde{x}_{kk}$$

where $x_{ij}, \tilde{x}_{ij}, y_{ij}, \tilde{y}_{ij}$, are respectively the elements of the matrices X, $X^{-1}$, Y and $Y^{-1}$.

The number of parameters for each speaker is (the covariance matrix is symetric):

$$parameters = \frac{P^2 + P}{2}$$

## 4. RESULTS

The results have been obtained with the following parameters:
- 49 bilingual speakers.
- One text (about 1 minute and the same text for all speakers) for computing the models.
- 5 sentences (the same for all speakers) for the test.
- No=number of bits of the codebook ranging from 0 to 7.
- Silence removal
- Frames of 240 samples with an overlap of 2/3.
- Hamming window and preemphasis of 0.95.

### 4.1 Vector quantization results

Table 1 summarizes the results for a vector quantization speaker identification method, with parameterizations LPCC-12, 16 and 20, and for codebooks ranging from 0 to 7 bits.

A new proposition for speaker modelling consists on the combination of two codebooks (one of Catalan and one of Spanish) of equal size, for obtaining a combined codebook of doubled size. Table 2 summarizes the results of this combined codebook.

| P  | t. | No=1 | No=2 | No=3 | No=4 | No=5 | No=6 | No=7 |
|----|----|------|------|------|------|------|------|------|
| 12 | S  | 41.6 | 32.2 | 53.9 | 89   | 96.7 | 97.6 | 99.2 |
| 12 | C  | 35.5 | 32.7 | 58.4 | 91.8 | 98.4 | 98.8 | 99.6 |
| 16 | S  | 46.5 | 44.1 | 60.8 | 91.4 | 96.7 | 98.4 | 99.6 |
| 16 | C  | 38.8 | 37.1 | 67.8 | 93.5 | 99.2 | 98.8 | 99.6 |
| 20 | S  | 49.8 | 44.9 | 64.1 | 91   | 96.7 | 98.4 | 99.6 |
| 20 | C  | 41.2 | 38.8 | 70.6 | 91.4 | 98.8 | 98.8 | 99.6 |

Table 2: Identification rates using a combined (S&C) VQ, t=test, S=Spanish, C=Catalan.

In order to study with more detail the differences between the Spanish and Catalan languages, the distortion quantizations of the recognized speakers has been accumulated for the whole speaker database in two ways:
- Taking into account the distortion between all the test sentences and all the codebooks.
-Taking into account only the distortions between the test sentences and the identified speaker.

Obviously, the values in the former case are greater, but the shape of the graphic is similar, so figure 2 presents the results for the latest case, as function of the number of bits of the codebook. X-X means training with language X and testing with language X (X=Spanish or Catalan), while cX means the combined codebook described previously, tested with language X.

From figure 2 it can be deducted that the vector distribution of the Catalan language can be better modelled than the Spanish one, with a given number of centroids in the codebook. It is interesting to observe that the quantization distortions with Catalan language are always smaller than the Spanish ones. This is valid for all the codebook sizes, and for all the studied values of P (12, 16 & 20).

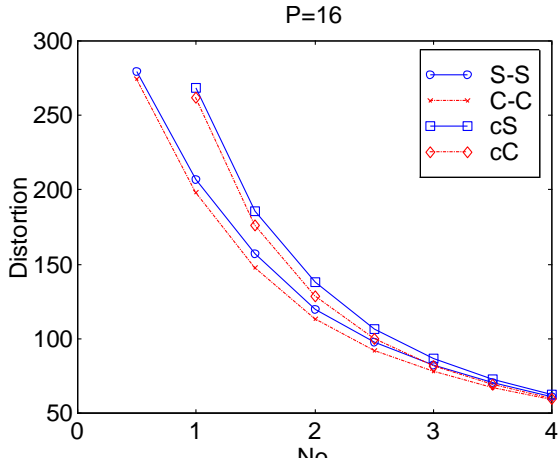

Figure 2: Quantization distortions

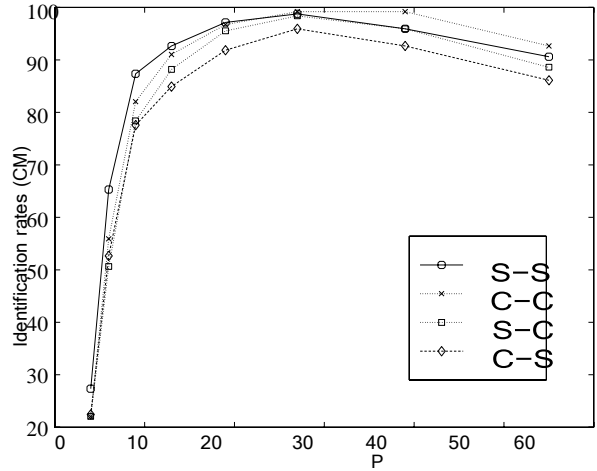

Figure 3: Identification rates using CM

Although it is not included for simplicity, we have observed that the quantization distortion for language X quantized with the codebook of language Y is very close to the distortion of language Y quantized with the codebook of language Y (that is, the distortion is mainly conditioned by the language used during training).

The identified speaker is chosen by the codebook that gives the minimal quantization distortion of the test sentence. The figure 2 represents the addition of all these distances.

**4.2 Covariance matrices**

The parameter that can be adjusted for modelling the speakers is the prediction order (P). That is, the dimension of the LPCC vectors.

We have studied several P values (table 3). It is important to see that a frame length of 240 samples is used, so for a correct LPC parameter estimation, the prediction order must not be higher than 24, because then the autocorrelation used in the Levinson-Durbin recursion cannot be properly estimated. For this reason, the recognition rates drop for high P values.

Another important fact is that a covariance matrix assumes that the modelled distribution is symmetrical. This assumption is not made in the VQ approach. Thus, for nonsymmetrical distributions the VQ approach could be more accurate.

For comparing both methods (quantization vectors and covariance matrices), we have used parameters Nq and P that require the same memory, as we see in table 4.

| Nq | Num. Parameters (aprox) | P |
|----|-------------------------|----|
| 0  | 12                      | 4  |
| 1  | 24                      | 6  |
| 2  | 48                      | 9  |
| 3  | 96                      | 13 |
| 4  | 192                     | 19 |
| 5  | 384                     | 27 |
| 6  | 768                     | 39 |
| 7  | 1536                    | 55 |

Table 4: Number of parameters used in VQ (P=12) and CM

## 5. CONCLUSIONS

This set of experiments has let us to obtain relevant conclusions about the relevance of language in speaker recognition. Main conclusions are:

- The Catalan database yields higher recognition rates than the Spanish one for a high number of parameters. Otherwise the Spanish language achieves better rates. We think that this is due to the higher number of vocalic phonemes (8 in Catalan against 5 in Spanish).

- With different test and train conditions there is a little decrease in recognition rate (about 1% for high resolution codebooks, and greater values for other models and methods)

- Although VQ achieves the highest recognition rates, the CM method is faster and in most cases requires less

parameters for modelling each speaker.

With this database we have made a study with different recognition systems (Vector quantization and Covariance matrices with an arithmetic-harmonic sphericity measure.), and several parameterizations. We also propose a method for improving the robustness of the recognition system combining both languages: a system with two models for each speaker (one model with each language).

## 6. FUTURE WORK

Our future work will include:
- The study of the influence of the variations when training and testing sesions are different.
- The robustness against different microphones.
- The robustness against noise.
- Performance with conversational speech.

| P | train/test | No=0 | No=1 | No=2 | No=3 | No=4 | No=5 | No=6 | No=7 |
|---|---|---|---|---|---|---|---|---|---|
| 12 | S-S | 63.3 | 69 | 76.3 | 89 | 95.9 | 97.1 | 99.2 | 98.8 |
| 12 | C-C | 58.4 | 66.9 | 74.3 | 87.3 | 96.3 | 98.4 | 99.2 | 99.6 |
| 12 | S-C | 49.4 | 62.4 | 64 | 82 | 90.6 | 94.7 | 96.3 | 97.6 |
| 12 | C-S | 47.3 | 59.2 | 69 | 88.2 | 92.7 | 94.7 | 96.7 | 98 |
| 16 | S-S | 67.3 | 72.2 | 78.8 | 90.2 | 96.3 | 98.4 | 98.4 | 98.4 |
| 16 | C-C | 62 | 72.2 | 74.3 | 86.5 | 98.8 | 99.2 | 99.2 | 100 |
| 16 | S-C | 55.9 | 65.7 | 65.7 | 82 | 91 | 95.9 | 98 | 98.8 |
| 16 | C-S | 52.7 | 61.6 | 70.6 | 87.8 | 95.9 | 97.6 | 98.8 | 98 |
| 20 | S-S | 68.6 | 71.8 | 78.8 | 94.3 | 96.7 | 98 | 98.8 | 99.2 |
| 20 | C-C | 64.9 | 74.3 | 79.6 | 90.6 | 98.8 | 99.2 | 99.6 | 100 |
| 20 | S-C | 57.1 | 70.2 | 69 | 86.1 | 93.5 | 96.7 | 98.4 | 99.2 |
| 20 | C-S | 54.3 | 62 | 71 | 91 | 97.1 | 96.3 | 98.8 | 98.4 |

Table 1: Identification rates using VQ (S=Spanish C=Catalan)

| train/test | P=4 | P=6 | P=9 | P=13 | P=19 | P=27 | P=39 | P=55 |
|---|---|---|---|---|---|---|---|---|
| S-S | 27.3 | 65.3 | 87.3 | 92.6 | 97.1 | 98.7 | 95.9 | 90.6 |
| C-C | 22 | 55.9 | 82 | 91 | 96.7 | 99.2 | 99.2 | 92.6 |
| S-C | 22 | 50.6 | 78.3 | 88.1 | 95.5 | 98.3 | 95.9 | 88.6 |
| C-S | 22.4 | 52.6 | 77.5 | 84.9 | 91.8 | 95.9 | 92.6 | 86.1 |

Table 3: Identification rates using CM (S=Spanish C=Catalan)